\documentclass[epj]{svjour}
\usepackage{latexsym}
\usepackage{graphics}
\usepackage{latexsym}                            
\usepackage{amsfonts}                            
\usepackage{amssymb}                             
\usepackage{amsmath}                             
\usepackage[mathscr]{eucal}                      

\newcommand{\ts}{\tau}                           

\begin{document}
\title{Helicity Dependent and Independent Generalized Parton Distributions of the Nucleon in Lattice QCD
\thanks{Based on talk given by Ph.H.} }
\author{Ph.~H\"agler\inst{1}  \and J.~W.~Negele\inst{2} \and D.~B.~Renner\inst{2,4}
 \and W.~Schroers\inst{2}
\and T.~Lippert\inst{3} \and K.~Schilling\inst{3} (LHPC Collaboration)
}                     
%
%
\institute{Department of Physics and Astronomy, De Boelelaan 1081, 1081 HV Amsterdam, The Netherlands 
\and Center for Theoretical Physics, Massachusetts Institute of Technology, Cambridge, MA 02139, USA
\and Department of Physics, University of Wuppertal, D-42097 Wuppertal, Germany 
\and current address: Department of Physics, University of Arizona, 1118 E 4th Street, Tucson, AZ 85721, USA }
\date{}
%
\abstract{
A complete description of the nucleon structure in terms of generalized parton distributions (GPDs) at twist 2 level
requires the measurement/computation of the eight functions $H$, $E$, $\widetilde{H}$,
$\widetilde{E}$, $H_{T}$, $E_{T}$, $\widetilde{H}_{T}$ and $\widetilde{E}
_{T}$, all depending on the three variables $x$, $\xi$ and $t$. In this talk, we present and discuss our first
steps in the framework of lattice QCD towards this enormous task. Dynamical lattice QCD results for the lowest three
Mellin moments of the helicity dependent and independent GPDs are shown in terms of their corresponding
generalized form factors. Implications for the transverse coordinate space structure of the
nucleon as well as the orbital angular momentum (OAM) contribution of quarks to the nucleon spin are discussed
in some detail.
} 

\authorrunning{Ph.H\"agler et al.}
\titlerunning{Helicity (In-) Dependent GPDs of the Nucleon in Lattice QCD}
\maketitle
\section{Introduction to GPDs}
\label{intro}
\begin{figure}
\begin{center}
\resizebox{0.40\textwidth}{!}
{
\rotatebox{270}{

  \includegraphics{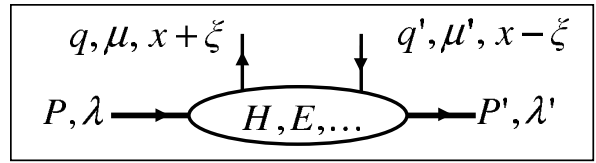}

}      }
\caption{Graphical representation of GPDs as part of a scattering amplitude}
\label{fig:GPDs}       
\vspace{-0.5cm}
\end{center}
\end{figure}

The generalized parton distributions (GPDs) of quarks in the nucleon we would like to compute
are defined through the parameterization of off-forward nucleon matrix elements
\begin{eqnarray}
  \label{VectorOp}
  &&\!\! \left\langle P',\lambda ' \right|
  \int \frac{d \lambda}{4 \pi} e^{i \lambda x}
  \bar \psi (-\frac{\lambda}{2}n)\!
  \gamma^\mu
  \psi(\frac{\lambda}{2} n)
  \left| P,\lambda \right \rangle   \nonumber \\
  &&\!\! \!\! \!\! \!\! =\overline U(P',\lambda ') \bigg( \!\gamma^\mu  H(x, \xi, t)
 + \frac{i \sigma^{\mu \nu} \Delta_\nu} {2 m} E(x, \xi, t) \!\bigg)   U(P,\lambda)
  ,
\end{eqnarray}
for the helicity independent respectively vector case, and
\begin{eqnarray}
  \label{AxialVectorOp}
  &&\!\! \left\langle P',\lambda ' \right|
  \int \frac{d \lambda}{4 \pi} e^{i \lambda x}
  \bar \psi (-\frac{\lambda}{2}n)\!
  \gamma_5 \gamma^\mu
  \psi(\frac{\lambda}{2} n)
  \left| P,\lambda \right \rangle   \nonumber \\
  &&\!\! \!\! \!\! \!\! =\overline U(P',\lambda ') \bigg(\! \gamma_5 \gamma^\mu  \tilde H(x, \xi, t)
 + \frac{\gamma_5 \Delta^\mu} {2 m} \tilde E(x, \xi, t) \!\bigg)   U(P,\lambda)
  ,
\end{eqnarray}
for the helicity dependent/axial vector case \cite{Muller:1998fv,Ji:1996nm,Radyushkin:1997ki}.
Here we dropped for simplicity the dependence on the
resolution scale $Q^2$ as well as the gauge links $U$ rendering the bilocal operators gauge invariant.
The definition of momenta and helicity labels can be inferred from Fig.(\ref{fig:GPDs}), and in addition
we note that $\Delta=P'-P$, $t=\Delta^2$, while $n$ is a light-cone vector with $n_\perp=0$, so that the
longitudinal momentum transfer is given by $\xi=-n\cdot \Delta/2$.
The proper definition of the tensor/quark helicity flip GPDs can be found in \cite{Diehl:2001pm}.
Some of the excitement about GPDs is due to their unifying nature, which can be read
off right away form the definitions Eqs.(\ref{VectorOp},\ref{AxialVectorOp}).
There is the forward limit, $\Delta \to 0$, in which $E$ and $\tilde E$ do not
contribute and the above definitions reduce exactly to the well known parton distributions,
$H(x, \xi=0, t=0)=q(x)=f_1(x)$, $\tilde H(x, \xi=0, t=0)=\Delta q(x)=g_1(x)$. Secondly,
by integrating over $x$, we recover the
defining equations for the the Dirac and Pauli form factors $F_1(t),F_2(t)$ as well as the axial
and pseudo-scalar form factors $g_A(t),g_P(t)$ (all to be taken with an appropriate flavor combination),
$\int dx H(x, \xi, t)=F_1(t)$, $\int dx E(x, \xi, t)=F_2(t)$  etc. Similar statements hold for the tensor
GPDs as well.
It is known for a long time that electromagnetic form factors encode information about the distribution
of charge in the nucleon in terms of e.g. charge radii, upon Fourier transformation in the momentum transfer $\Delta$
to coordinate or impact parameter space, $b$. The obvious question is if and how this
translates to GPDs. Due to the Heisenberg uncertainty relation, it seems at first sight that
the GPDs as functions of the quark longitudinal momentum fraction $x$ and $\Delta$ will not allow for a
probability interpretation, since longitudinal momentum and position in longitudinal direction cannot be
determined simultaneously. As M. Burkardt has pinpointed\cite{Burkardt:2000za}, 
it turns out that just by integrating over
longitudinal position, that is setting $\xi=0$, ones gets functions of $x$ and $b_\perp$ which have
a perfect probability density interpretation. To be precise, it is shown that 
\begin{eqnarray}
q(x,b_\perp) &\equiv& H(x,0,b_\perp)\nonumber\\
&\equiv&\int d^2\Delta_\perp e^{-i b_\perp \cdot \Delta_\perp} H(x,\xi=0,t=-\Delta_\perp^2)
\end{eqnarray}
is positive definite and interpreted as helicity independent quark probability density in $b_\perp$ (and $x$),
even in a relativistic framework. Similar statements hold true for the impact parameter dependent
helicity and tensor/transverse spin distributions, $\Delta q(x,b_\perp)$ and $\delta q(x,b_\perp)$.
It is quite illuminating to rewrite $q(x,b_\perp)$ using a fock state decomposition and
nucleon wave functions $\Psi _{n,c}$
(overlap representation)\cite{Diehl:2000xz,Brodsky:2000xy,Diehl:2002he},
\begin{eqnarray}
&&q(x,b_{\perp }) =\sum\limits_{n,c}\sum\limits_{a}(4\pi)^{n-1}\int
\left[ \prod\limits_{j=1}^{n}dx_{j} d^{2}r_{\perp j}\right]
\nonumber\\
&&\times\delta \left(
1-\sum\limits_{j=1}^{n}x_{j}\right)
 \delta ^{2}\left(
\sum\limits_{j=1}^{n}x_{j}r_{\perp j}\right)
 \delta \left( x-x_{a}\right)
\nonumber\\
&&\times \delta ^{2}\left( b_{\perp
}-(1-x)r_{\perp a}+\sum\limits_{j\neq a}^{n}x_{j}r_{\perp j}\right)
\nonumber\\
&&\times \left|\Psi _{n,c}(x_{1},\ldots ;r_{\perp 1},\ldots)\right|^2 \,,
\label{Fock}
\end{eqnarray}
where $n$ is the number of partons in a given Fock state and the sums are carried out over
all contributing $n$-parton configurations; $a$ denotes the active parton. Eq.(\ref{Fock})
shows clearly the anticipated positivity of the impact parameter distribution. Furthermore it
becomes clear that detailed knowledge of the GPDs will certainly lead to an improved understanding
of the all-important nucleon wave function. The first two delta functions in Eq.(\ref{Fock}) ensure
conservation of longitudinal and transverse momentum respectively that the center of momentum (CM) is
$R_\perp=\sum_{j=1}^{n}x_{j}r_{\perp j}=0$. The last delta function shows on the one hand that $b_\perp$ is
the distance of the active parton to the CM, $b_\perp=r_{\perp a}-\sum_{j}^{n}x_{j}r_{\perp j}=r_{\perp a}$.
On the other hand, written as in Eq.(\ref{Fock}), we see explicitly what happens if $x$ is close to one
(while $x_{j\neq a}\to 0$).
Taking into account that the distribution will finally vanish in the limit $x\to 1$,
we observe that the normalized distribution
becomes strongly peaked at the origin, 
$\lim_{x \to 1}q(x,b_{\perp })/q(x) \varpropto \delta ^{2}\left( b_{\perp}\right)$.
In momentum space, we get the remarkable prediction that the normalized GPD $H(x,0,t)/q(x)$ will
be constant in the limit of $x\to 1$\cite{Burkardt:2002hr,Burkardt:2004bv}.

In addition to what was mentioned before, GPDs offer probably for the first time an undisputed way to learn
about the contribution of quark orbital angular momentum (OAM) to the nucleon spin. Starting from the
QCD energy momentum tensor and applying the standard Noether procedure, we get the following
conservation laws (the first one also frequently called nucleon spin sum rule)\cite{Ji:1996ek},
\begin{eqnarray}
\label{spinsumrule}
\frac{1}{2} &=& J^q + J^g=\frac{1}{2}\left(A_{20}(0)+B_{20}(0)\right)^{q+g}\,,\\
1 &=& A_{20}^{q+g}(0),\,\,
0 = B_{20}^{q+g}(0)\,.
\label{sumrules}
\end{eqnarray}
At this point it is essential to note that $A_{20}^{q,g}(0)$ is equal to the momentum fraction
$\langle x\rangle^{q,g}$
while $B_{20}^{q,g}(0)$ is the second moment of the GPD $E$, $B_{20}^{q,g}(0)=\int dx x E^{q,g}(x,0,0)$.
The vanishing of the so called anomalous gravitomagnetic moment
$B_{20}^{q+g}(0)$ has also been derived in the light-front formalism\cite{Brodsky:2000ii}.
Given the two Eqs.(\ref{sumrules}), the nucleon spin sum rule Eq.(\ref{spinsumrule})
seems to be a trivial identity.
This is only true as long as we do not decompose into quark and gluon as well as spin and orbital degrees
of freedom. A standard way of rewriting the spin sum rule is \cite{Ji:1996ek}
\begin{equation}
\label{spinsumrule2}
\frac{1}{2} = \frac{1}{2}\Delta q + L^q + J^g=\frac{1}{2}\left(A_{20}(0)+B_{20}(0)\right)^{q+g}\,,
\end{equation}
where we did not further split the gluon piece into spin and orbital angular momentum, since
it is not clear how to define $L^g$ properly on an operator level.
Assuming that a commonly accepted and reasonable definition for the gluon spin $\Delta g$ exists,
we may however just set $L^g=J^g-\Delta g$\cite{Ji:1996ek}.
This ambiguity does not affect the following discussions,
since our current nucleon structure lattice computations are restricted to quark observables anyway.
From Eq.(\ref{spinsumrule2}) we find that quark OAM in the nucleon can be computed once the quark momentum fraction
$\langle x\rangle^q$, the quark spin $\Delta q$
and the second moment of the GPD $E^q(x,0,0)$ are known.

Let us now comment briefly on the relation of the magnetic moment $\mu$ respectively $F_2(0)$ to quark OAM.
It has been pointed out frequently \cite{Jain:1993jf}
that a non-vanishing $F_2(0)$
requires necessarily the change of orbital
quantum numbers $\Delta l=\pm 1$ in the corresponding wave functions. It is tempting to assume
that for this reason a large proton (flavor combination $2/3u-1/3d$) magnetic moment
$\mu^p=1+F_{2}^p(0)=1+\int dx E^p(x,0,0)$ implies a substantial quark OAM contribution to the nucleon
spin. This is not the case. The relevant flavor combination for the spin sum rule
is of course $u+d$, and assuming iso-spin symmetry, the experimental value of the iso-singlet $F_{2}^{u+d}(0)$
is rather small and negative $\approx -0.36$ due to the fact that the separate flavors are
of almost equal magnitude but opposite sign. In addition, the value for the second moment
$B_{20}^{u+d}(0)$ is probably even smaller by a factor $\approx 2\ldots 4$ or more \cite{Diehl:2004cx}.
Hence the \emph{total} quark OAM in the sum rule Eq.(\ref{spinsumrule2}) receives little
contribution from the GPD $E$ \cite{Hoodbhoy:1998yb}. Even for the separate flavors
$u$ and $d$, it is by no means clear that relatively large values of $B^{u,d}_{20}$ lead to
large $L^{u,d}$, since there may be cancellations of $\langle x\rangle^{u,d}$, $\Delta q^{u,d}$ and
$B^{u,d}_{20}$ in $L^q=1/2 (\langle x\rangle +B_{20}-\Delta q)$.

In summary, the large proton magnetic moment is not related
to the size of the quark OAM contribution $L_q$ to the nucleon spin and therefore
of no help with regard to the "nucleon spin crisis". This does not contradict the fact that we
need a change of orbital quantum numbers in the wave function to get a non-zero $F^p_2(0)$.

\section{Outline of Lattice Computation of GPDs}
\label{sec:1}
Since it is not feasible to evaluate matrix elements of bilocal light-cone operators directly on the lattice,
the first step is to transform the LHS in Eqs.(\ref{VectorOp},\ref{AxialVectorOp})
to Mellin space by integrating over $x$, i.e. $\int dx x^{n-1}$. The resulting matrix elements
of towers of local operators are in turn parameterized in terms of generalized
form factors (GFFs), here generically denoted by $A(t),B(t)\ldots$,
\begin{eqnarray}
&&\!\!\!\!\left\langle P',\lambda '\right|{\cal
    O}_\Gamma \!\left|P,\lambda\right\rangle=\left\langle P^{\prime },\lambda '\right| \bar{\psi}(0)
    \Gamma iD^{\{\mu _{1}}\!\!\cdots iD^{\mu _{n}\}}\psi (0)\left| P, \lambda \right\rangle \nonumber \\
 &&\!\!=\overline{U}(P',\lambda ')\left(a_{\Gamma }^{\mu
_{1}\ldots \mu _{n}}A(t)+b_{\Gamma }^{\mu
_{1}\ldots \mu _{n}}B(t)+\cdots \right)U(P,\lambda),
\label{ME1}
\end{eqnarray}
where we do not explicitly show the subtraction of traces.
The explicit parameterization for the tower of vector operators, $\Gamma\hat=\gamma^\mu$, is
presented in \cite{Hoodbhoy:1998yb} in
terms of the $(2n+1)$ independent GFFs $A_{ni}(t),B_{ni}(t)$ and $C_{n0}(t)$.
In the axial vector case, the corresponding expression in terms of the
$\left(2\left\lfloor \frac{n}{2}\right\rfloor +2\right)$ GFFs
$\tilde A_{ni}(t)$ and $\tilde B_{ni}(t)$ is shown in \cite{Diehl:2003ny}. Recently, parameterizations for the
tensor GPDs became available as well \cite{Hagler:2004yt,Chen:2004cg},
allowing for a first-time computation of these objects in
lattice QCD.
The GFFs which parameterize the matrix elements depend only on the invariant momentum transfer squared $t$.
Therefore, having knowledge only of the GFFs, the directional dependence via $\xi=-n\cdot \Delta/2$
seems to be lost and there is no immediate way to tell where the information on
the transverse or longitudinal structure is hidden.
It turns out, however, that the moments of the GPDs and the GFFs are related by comparatively
simple polynomial relations in $\xi$ which restore the full directional dependence \cite{Ji:1997gm}.
For the vector operator we have for example that
\begin{eqnarray}
H_{n}(\xi ,t) &\equiv& \int\limits_{-1}^{1}dxx^{n-1}H(x,\xi ,t)
=\!\!\!\!\sum_{i=0, \text{even}}^{n-1}\left( -2\xi \right) ^{i}A_{ni}(\Delta
^{2})\nonumber\\
&&+\left. \left( -2\xi \right) ^{n}C_{n0}(\Delta ^{2})\right| _{n
\text{ even}}  \nonumber \\
E_{n}(\xi ,t) &=&\sum_{i=0,
 \text{even}}^{n-1}\left( -2\xi
\right) ^{i}B_{ni}(\Delta ^{2})\nonumber\\
&&-\left. \left( -2\xi \right)
^{n}C_{n0}(\Delta ^{2})\right| _{n\text{ even}}.  \label{invv}
\end{eqnarray}
showing that for a purely transverse momentum transfer, $\xi=0, t=-\Delta_\perp^2$,
the complete nucleon structure concerning $H$ is
represented by the set of GFFs $A_{n,0}(t)$, $n=1\ldots\infty$. The well known form factors
$F_1=A_{10}$, $g_A=\tilde A_{10}$ and $g_T=\tilde A_{T10}$
give in this respect just transverse distributions of quarks in the nucleon, integrated
over quark longitudinal momentum $x$.

On the lattice side, we have to study and eventually compute nucleon two- and three-point correlation functions.
The three-point function is defined by
\begin{equation}
  C_{\cal O}^{\text{3pt}}(\tau,P',P) = \sum_{j,k}
    \tilde\Gamma_{jk}\left\langle N_{k}
    (\tau_{\text{snk}},P'){\cal O}_\Gamma(\tau)
  \overline{N}_{j}(\tau_{\text{src}},P)\right\rangle,
  \label{threept}
\end{equation}
where $\tilde \Gamma$ is a (spin-)projection matrix and the operators $N$ and $\overline{N}$ create respectively destroy states with the quantum
numbers of the nucleon, and ${\cal O}_\Gamma$ stands in general for the tower of local operators we would like to investigate,
inserted at a certain timeslice $\tau$.
To see the relation of $C_{\cal O}$  to the parameterization Eq.(\ref{ME1}) respectively the GFFs most clearly,
let us rewrite Eq.(\ref{threept}) using complete sets of states and the time evolution operator,
\begin{eqnarray}
  \label{threept2}
  &&C_{\cal O}^{\text{3pt}}(\ts,P',P) =
  N(P',P)e^{-E(P)(\ts-\ts_{\text{src}})-E(P')(\ts_{\text{snk}}-\ts)}
   \nonumber\\
   &&\times \sum_{\lambda,\lambda '}
    \left\langle P',\lambda '\right|{\cal
    O}_\Gamma \left|P,\lambda \right\rangle \overline{U}(P,\lambda)\tilde\Gamma U(P',\lambda ')+\ldots
\end{eqnarray}
and $N(P',P)$ denotes some normalization factors. The dots in (\ref{threept2}) stay for excited states
with energies $E'>E$ which are exponentially suppressed as long as $\ts-\ts_{\text{src}}>>1/E', \ts_{\text{src}}-\ts>>1/E'$.
Inserting Eq.(\ref{ME1}), we can carry out the sums over polarizations and get
\begin{eqnarray}
  \label{threept3}
  &&C_{\cal O}^{\text{3pt}\mu_{1}\ldots \mu _{n}}(\ts,P',P) \varpropto
   \text{Tr}\bigg[ \tilde\Gamma(\not{\!P}^{\prime}+m)
   \nonumber\\
   && \times (a_{\Gamma }^{\mu_{1}\ldots \mu _{n}}A(t)+b_{\Gamma }^{\mu_{1}\ldots \mu _{n}}B(t)+\cdots )
   (\not{\!P}+m)
  \bigg]\,,
\end{eqnarray}
where we dropped for simplicity all pre-factors but showed the dependence of the three-point function
on the indices $(\mu_{1}\ldots \mu _{n})$. The trace in Eq.(\ref{threept3}) can be explicitly evaluated, while
the normalization factor and the exponentials in Eq.(\ref{threept2}) will be canceled out by constructing an
appropriate $\tau$-independent ratio $R$ (see e.g. \cite{Hagler:2003jd}) of the two- and three-point functions.
On the lattice side, the ratio $R$ is numerically evaluated and then equated with the corresponding sum
of GFFs times $P$ and $P'$-dependent calculable pre-factors, coming from the trace in Eq.(\ref{threept3}). For a given
moment $n$, this is done simultaneously for all contributing index- combinations $(\mu_{1}\ldots \mu _{n})$
and all lattice momenta $P,P'$ corresponding to the same value of $t$.
Following this, we end up with an in general overdetermined set of equations from which we finally extract the GFFs.
Many details related to the actual calculations can be found in \cite{Hagler:2003jd}.
\begin{figure}
\resizebox{0.4\textwidth}{!}
{
  \includegraphics{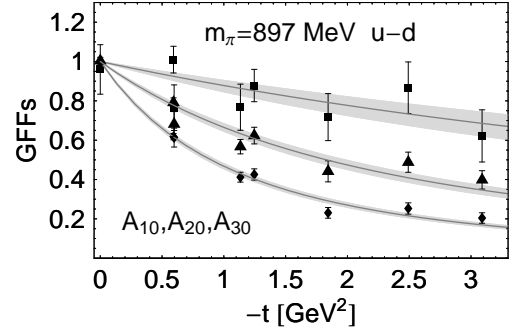}
}
\caption{The lowest three moments of $H(x,\xi=0,t)$, normalized to $A^{\text{dipole}}_{n0}(0)=1$}
\label{fig:1}       
\end{figure}
\section{Numerical results}
\label{sec:2}
\begin{figure}
\resizebox{0.4\textwidth}{!}
{
  \includegraphics{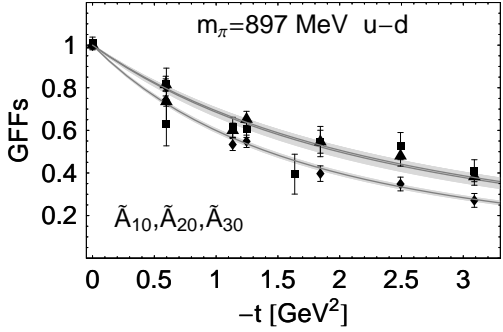}
}
\caption{The lowest three moments of $\tilde H(x,\xi=0,t)$, normalized to $\tilde A^{\text{dipole}}_{n0}(0)=1$}
\label{fig:2}       
\end{figure}

\begin{figure}
\resizebox{0.4\textwidth}{!}
{
  \includegraphics{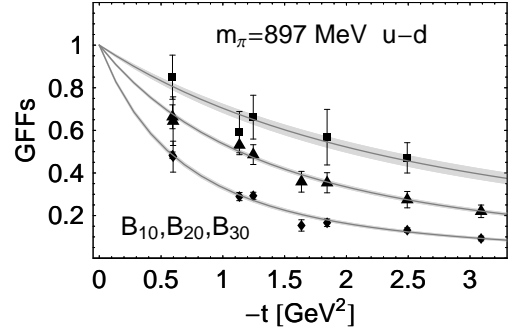}
}
\caption{The lowest three moments of $E(x,\xi=0,t)$, normalized to $B^{\text{dipole}}_{n0}(0)=1$}
\label{fig:3}       
\end{figure}

We use for our computations $O(200)$ SESAM configurations with rather heavy (unimproved) Wilson fermions
at a coupling of $\beta=5.6$.
Although our results are based on a dynamical QCD calculation (including fermion loops),
we do not take into account the so-called disconnected contributions which are numerically much
harder to evaluate. We therefore prefer to show the iso-vector channel $u-d$,
for which the disconnected pieces, due to iso-spin symmetry, would cancel out anyway.
The lattice size is $16^3\times 32$, and the physical lattice extension is $L\approx 1.6\text{fm}$. The results shown
below corresponds to a pion mass of $m_\pi\approx 897\text{MeV}$ ($\kappa=0.1560$), equivalent to a pion Compton
wavelength of $\approx 1/7$ of the lattice extension. The charge radius of the proton turns out to
be $\approx 0.5\text{fm}$ for these lattice parameters. A selection of our results on the
$t$-dependence of the GFFs is shown in Figs.(\ref{fig:1}-\ref{fig:3}) \cite{Schroers:2003mf,Hagler:2003is}.
In all cases we used as a lowest order approximation a dipole to fit the lattice numbers.
The values for the corresponding dipole masses are
\begin{eqnarray}
m_{D}\left(A_{1\ldots 3,0}\right) = 1.47\!\pm\! .03,2.10\!\pm\! .08,3.86\!\pm\! .49 \text{ GeV}\nonumber\\
m_{D}\left(\tilde A_{1\ldots 3,0}\right) = 1.85\!\pm\! .03,2.22\!\pm\! .06,2.23\!\pm\! .09 \text{ GeV}\nonumber\\
m_{D}\left(B_{1\ldots 3,0}\right) = 1.16\!\pm\! .02,1.66\!\pm\! .02,2.28\!\pm\! .09 \text{ GeV}\nonumber\\
\end{eqnarray}
It is astonishing how strong the slope in $t$ decreases, 
going from the first to the third moment of $H$, Fig.(\ref{fig:1}).
But this is what we generically expected from the first-principles
discussion below Eq.(\ref{Fock}): The higher the moment
$n$, the more weight lies on values of $x$ close to one. 
In the extreme case, $\lim_{n\to\infty} A_{n0}(t)/A_{n0}(0)=1$,
independent of $t$. Since the average value of $x$ of the
distribution $q^{u-d}(x)=H^{u-d}(x,0,0)$ is $\approx 0.22$
and changes for $x^{2}q^{u-d}(x)$ (third moment) just to $\approx 0.4$ \cite{Dolgov:2002zm},
the effect appears to set in quite promptly \cite{Negele:2004iu}.
As Fig.(\ref{fig:2}) indicates, the picture looks different for the helicity-dependent GPD $\tilde H$. We still see
a change in the dipole mass going from $\tilde A_{10}$ to $\tilde A_{30}$, but it is much less pronounced
than in the helicity averaged case. If this picture stands up to further studies, one could draw
significant conclusions about the helicity and transverse momentum structure of the
nucleon wave function from this observation.
The physical interpretation of $E$ and the corresponding GFFs $B_{ni}$ concerning the transverse nucleon structure
is more involved \cite{Burkardt:2002hr}.
Still, it is interesting to note that from the dipole fit, we get a noticeable change in the slope in $t$
for the lowest three moments as well, Fig.(\ref{fig:3}). The lattice results for the helicity dependent
analogue, $\tilde E$, suffer at least for the higher moments from strong fluctuations and are not shown in this work.

\begin{figure}
\resizebox{0.4\textwidth}{!}
{
  \includegraphics{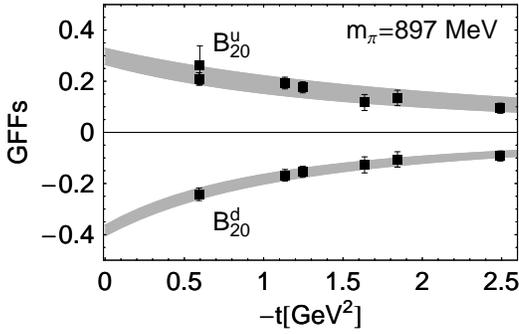}
}
\caption{$B^{u}_{20}$ and $B^{d}_{20}$ with dipole fits}
\label{fig:4}       
\end{figure}

Concerning quark OAM, we have plotted in Fig.(\ref{fig:4}) $B^{u}_{20}$ and $B^{d}_{20}$ separately. Note that
in this case the disconnected pieces have been neglected, leading to a systematic error 
of unknown size.
The dipole fits give
\begin{equation}
\label{B20}
B^{u,\text{dipole}}_{20}(0)=0.29\!\pm\! .03,\,\,\, B^{d,\text{dipole}}_{20}=-0.38\!\pm\! .02
\end{equation}
This is exactly what we expected from the discussion at the end of section (\ref{intro}). The up and down
contributions are similar in size but opposite in sign, giving a small and negative iso-singlet
$B^{u+d}_{20}(0)=-0.09\pm 0.04$, almost compatible with zero. Using the results for the momentum fraction
and the quark spin contribution \cite{Dolgov:2002zm}, we get for quark OAM,
$L_q=1/2 (\langle x\rangle+B_{20}-\Delta q)$,
\begin{equation}
\label{Lud}
L^{u}=-0.088\!\pm\! 0.019,\,\,\, L^{d}=0.036\!\pm\! 0.013\,.
\end{equation}
Both flavors separately give a rather small contribution of the order of $17\%$ ($7\%$)
for u-quarks (d-quarks) to the nucleon spin,
indeed due to cancellations in quark momentum fraction, spin and $B_{20}$.
Adding $u$ and $d$ contributions gives a very small and negative total OAM.
We would like to remind the reader that our calculation correspond to a situation
where the pion weighs $897\text{ MeV}$.
The corresponding result for the iso-vector magnetic moment,
for example, is $\approx 20\%$ and the quark momentum fraction $\approx 40\%$ above
the experimental value at this pion mass, so that the numbers have
to be taken with appropriate caution. A detailed comparison with experimental results
calls for lattice computations at much lower pion masses and an appropriate chiral extrapolation
(work in progress, see e.g. \cite{talks}).

Lattice calculations of the two lowest moments of the helicity independent
GPDs in quenched QCD have also been presented in \cite{lattice3}.
Calculations in the quenched approximation of quark angular momentum including disconnected
contributions have been reported in \cite{Mathur:1999uf,Gadiyak:2001fe}.


\section{Summary}
\label{summary}
The impact of the observation that GPDs are directly related to coordinate space distributions of
partons in the nucleon is profound and even led to the notion of
"femto-photography" and "tomographic images" of the nucleon \cite{Ralston:2001xs,Burkardt:2002ks}.
Certainly a lot of effort still has to be put into the study of
the impact parameter dependence of GPDs in order to get a detailed 
picture of the 3-dimensional nucleon structure.
Acknowledging that it is up to now hard to come by
experimental results on GPDs via deeply virtual Compton scattering and similar setups,
interesting analyses making use of experimental data on form factors and
structure functions to describe GPDs have been published recently \cite{Diehl:2004cx}
(see also \cite{Tiburzi:2004mh}).
As we have shown, however, lattice QCD as a first principle technique plays already now an important role
in unraveling main qualitative features of GPDs and thereby of the nucleon wave function.
As soon as the infrastructure allows for extensive calculations using chiral fermions, we expect
that the methods presented here will give an essential contribution to the quantitative understanding of the nucleon
structure in terms of GPDs. Promising attempts in this direction are under way
in the framework of a hybrid calculation using Astaq sea quarks and domain-wall valence quarks \cite{talks}.

%
%
%
%
%

\subsection*{Acknowledgments}

Ph.H. would like to thank S. Brodsky, M. Diehl and J.M. Zanotti for helpful discussions.
W.S.~is supported by the Feodor-Lynen program 
of the Alexander von Humboldt foundation.
This research has been supported by the EU Integrated Infrastructure Initiative
"Study of Strongly Interactive Matter" under contract number RII3-CT-2004-506078
and by the U.S.~Department of Energy (D.O.E.) under cooperative research
agreement \#DF-FC02-94ER40818.


\end{document}